\documentclass{lmcs}
\pdfoutput=1

\usepackage{lastpage}
\lmcsdoi{20}{3}{20}
\lmcsheading{}{\pageref{LastPage}}{}{}%
{May~17,~2023}{Sep.~04,~2024}{}

\usepackage[utf8]{inputenc}

\usepackage{graphicx,amsmath,amsthm,latexsym,amssymb,color,tikz,xspace}
\usepackage[hidelinks]{hyperref}
\usepackage{mathtools}
\usepackage{url}

\newtheorem{question}[thm]{Question}

\newcommand{\antidom}{\mathop{\sim}}
\newcommand{\TRA}{\ensuremath{\mathbb{TRA}}\xspace}
\newcommand{\BRA}{\ensuremath{\mathbb{BRA}}\xspace}
\newcommand{\FA}{\ensuremath{\mathbb{FA}}\xspace}
\newcommand{\FO}{\textrm{\upshape FO}\xspace}
\newcommand{\GSO}{\textrm{\upshape GSO}\xspace}

\newcommand{\dom}{\operatorname{dom}}
\newcommand{\id}{\operatorname{id}}
\newcommand{\D}{\operatorname{\sf D}}
\newcommand{\R}{\operatorname{\sf R}}
\newcommand{\iunion}{\mathbin{\mathrlap{\sqcup}{^{\,1}}}}

\usepackage[]{hyperref}
\hypersetup{
    colorlinks=true,
}

\begin{document}

\title{Preservation theorems for Tarski's relation algebra}

\author[B.~Bogaerts]{Bart Bogaerts\lmcsorcid{0000-0003-3460-4251}}[a]
\author[B.~ten Cate]{Balder ten Cate\lmcsorcid{0000-0002-2538-5846}}[b]
\author[B.~McLean]{Brett McLean\lmcsorcid{0000-0003-2368-8357
}}[c]
\author[J.~Van den Bussche]{Jan Van den Bussche\lmcsorcid{0000-0003-0072-3252}}[d]

\address{Vrije Universiteit Brussel, Brussel, Belgium}	
\email{bart.bogaerts@vub.be}  

\address{ILLC, University of Amsterdam,
Amsterdam, The Netherlands}	
\email{b.d.tencate@uva.nl}  

\address{Ghent University, Ghent, Belgium}
\email{brett.mclean@ugent.be}

\address{Hasselt University, Hasselt, Belgium}
\email{jan.vandenbussche@uhasselt.be}

\newcommand{\diagonaltablehead}[1]{%
  \phantom{nn}%
  \llap{\rotatebox[origin=r]{-45}{#1}}%
}

\begin{abstract}
We investigate a number of semantically defined fragments
of Tarski's algebra of binary relations, including the
function-preserving fragment. We address the 
question of whether they are generated by a finite set of 
operations. We obtain several positive and negative results
along these lines. Specifically, the homomorphism-safe fragment is finitely generated (both over finite and over arbitrary structures). The function-preserving fragment is not finitely generated (and, in fact, not expressible by any finite set of guarded second-order definable function-preserving operations). Similarly, the total-function-preserving fragment is not finitely generated (and, in fact, not expressible by any finite set of guarded second-order definable total-function-preserving operations). In contrast, the forward-looking function-preserving fragment is finitely generated by composition, intersection, antidomain, and preferential union. Similarly, the forward-and-backward-looking injective-function-preserving fragment is finitely generated by composition, intersection, antidomain, inverse, and an `injective union' operation.
\end{abstract}

\maketitle

\section{Introduction}

\begin{table}[t]
    \centering
    \begin{tabular}{llllll}
        && \diagonaltablehead{homomorphism-safe} & 
        \diagonaltablehead{$\subseteq$-safe} &
        \diagonaltablehead{function-preserving} & 
        \diagonaltablehead{forward} \\
    \hline
    id                & identity relation & yes & yes & yes & yes \\
    $\emptyset$       & empty relation    & yes & yes & yes & yes \\
    $\top$       & universal relation (all pairs)    & yes & yes & no & no \\[2mm]
    $-(\,\cdot\,)$        & complement        & no  & no & no & no \\
    $(\,\cdot\,)^\smile$    & inverse         & yes & yes & no & no \\
    $\D(\,\cdot\,)$ & domain ($\D(R) = \{(x,x)\mid  R(x,y)\}$) 
                                          & yes & yes & yes & yes \\
    $\R(\,\cdot\,)$ & range ($\R(R) = \{(y,y)\mid R(x,y)\}$) 
                                          & yes & yes & yes & no \\
    $\antidom(\,\cdot\,)$ & antidomain ($\antidom R = \{(x,x)\mid \neg\exists y\, R(x,y)\}$)              & no & no & yes & yes \\[2mm]
    $\cdot\cup\cdot$    & union           & yes & yes & no & yes \\ 
    $\cdot\cap\cdot$    & intersection    & yes & yes & yes & yes \\
    $\cdot\setminus\cdot$ & relative complement & no & yes & yes & yes \\
    $\cdot\circ\cdot$   & composition     & yes & yes & yes & yes \\
        $\cdot\ltimes\cdot$ & left semi-join ($R\ltimes S=\{(x,y)\in R\mid \exists z\,S(y,z)\}$)           & yes & yes & yes & yes \\
    $\cdot \sqcup \cdot$ & preferential union ($R\sqcup S = R \cup \{(x,y)\in S\mid \neg\exists z\, R(x,z)\}$) & no & no & yes & yes
    \end{tabular}
    \caption{Operations on binary relations}
    \label{tab:operations}
\end{table}

Just as Boolean algebra can be viewed as a language for describing operations on sets,
\emph{Tarski's relation algebra} (\TRA) is a language for describing operations on binary relations.
It consists of a small, finite collection of
 operations on binary relations (which includes, for instance,
composition and union), governed by natural equations such as
$R\circ (S\cup T) = (R\circ S) \cup (R\circ T)$. 
The origins of \TRA trace back to the 19th century, and, more specifically, to the work of  Augustus De Morgan and Charles Peirce, but its study intensified 
when it was picked up by Tarski and his students in the 1940s
\cite{Tarski41:RA,maddux_originra,pratt_relcalc}.
We can view \TRA
as a language for specifying operations on binary relations. Its expressive power, in terms of the term-definable
operations, corresponds precisely to the three-variable fragment
of first-order logic ($\FO^3$) \cite{Tarski1987:formalization}.

Many modern graph and tree query languages, such as regular path queries, SPARQL, and XPath, which describe ways of navigating through graph-structured data, can be identified with  variants of \TRA, each involving a different set of allowed operations.
This has 
generated an interest in systematically understanding the
expressive power of fragments and extensions of
\TRA~\cite{rafragments_ins,navtc,Hellings2022}.

Here, we study the question whether certain
semantically-defined fragments of \TRA can be generated by a finite set of operations.
One known positive result along these lines is the following, where $\BRA(\mathcal{O})$ denotes the binary relation algebra
generated by the operations in $\mathcal{O}$ 
(see~Table~\ref{tab:operations} for a definition of the operations).

\begin{thmC}[\cite{vBenthem1998:bisimulation}]
\label{thm:bis-safe}
    A \TRA-term is ``bisimulation safe'' if and only if it is equivalent
    to a $\BRA({\id, \circ, \cup, \sim})$-term.
\end{thmC}

The precise definition of \emph{bisimulation} and of \emph{bisimulation safety} is not important for us here. It suffices
that bisimulation is an important equivalence relation
that captures behavioral equivalence of processes,
and that an operation on binary relations is bisimulation safe if commutes, in a natural way, with  bisimulation. 

We can think of Theorem~\ref{thm:bis-safe} result as analogous to a \emph{preservation theorem} in model theory: 
it correlates a semantic property with expressibility
in a natural, finitely-generated, syntactic fragment. The above result may suggest that various other semantically-defined fragments of \TRA could be similarly characterised syntactically by a finite basis of operations.
One particular prominent semantic fragment that arises naturally in different contexts, is the function-preserving
fragment of \TRA \cite{mclean-thesis}. An operation 
on binary relations is
said to be \emph{function-preserving} if, 
whenever the input relations are partial functions,
so is the output relation.
It is a natural question, and an open problem in the community (although we could not locate an explicit reference) whether the
function-preserving
fragment of \TRA is finitely generated. 

\paragraph{Contributions}
As our main contribution, we establish the following positive and negative results:
\begin{itemize} 
\item 
The \emph{homomorphism-safe} fragment of \TRA
is finitely generated (Section~\ref{sec:monotone}).
\item
   The \emph{function-preserving} fragment of \TRA
is not finitely generated  (and, in fact, not expressible by any finite set of guarded second-order definable function-preserving operations). The same holds for the \emph{total-function-preserving} fragment
(Section~\ref{sec:func-pres}).
\item 
The \emph{forward function-preserving} fragment and the \emph{local injective-function-preserving} fragment
are finitely generated (Section~\ref{sec:func-pres-positive}).
\end{itemize}
We study each of these fragments both in the general case (i.e., where the input relations may be relations over an infinite domain)  and in the finite.

Naturally, there are many other semantic fragments of \TRA for which one could ask the same finite-generatedness question. Our
intention, with the above results, is to 
provide a sample of interesting results when it comes to 
the question of finite generation for semantic fragments of \TRA. In the concluding Section~\ref{sec:conclusion}, we will further comment on directions for future work and connections to the formalisms we mentioned in our motivation above.

\paragraph{Related Work}

B\"orner and P\"oschel~\cite{Borner1991:clones} studied whether various clones of operations on binary relations \emph{over a fixed finite structure} are finitely generated. Their study includes the ``logical clone'' (which is the set of all first-order definable operations) as well as the ``positive clone'' (which is the set of all operations
definable by positive-existential first-order formulas). 
Our investigation is different in that we are
interested in the existence of finite bases \emph{over all (finite) structures}. 
We will further comment on the relationship between our results and those
by B\"orner and P\"oschel in Section~\ref{sec:monotone}. 

Andr\'eka et al.\ \cite{Andreka1985:clones} and B\"orner \cite{Borner1986:one}
consider the problem whether certain finitely generated clones of operations on binary relations are in fact generated by a single operator (analogous to the Sheffer stroke in Boolean algebra), and what is the minimum possible arity of 
such an operation.

There is a substantial literature on algebras of partial functions (that is, 
function-preserving fragments of \TRA), focusing on the axiomatisation of their first-order theories as well as computational aspects such as 
decidability and the finite model property.
An in-depth overview of known results along these lines can be found in~\cite{mclean-thesis}. 

In the literature on temporal logics, there have been extensive studies concerning the existence of temporal logics generated by a finite set of operations, that are expressively complete for first-order logic
in the sense of Kamp's theorem~\cite{Kamp1968} (see~\cite{GHR:book} for an overview). One of the main differences with our setting is that, in temporal logic, the operators are typically
monadic (i.e., they correspond to \FO-formulas in one free variable), whereas in our case, the operators act on, and produce, binary relations (and hence correspond to \FO-formulas in two free variables). Closer to our setting is Venema~\cite{Ven90}, who studies expressive completeness for interval temporal logics, and showed that, on dense linear orders, 
no finite set of binary operations is expressively complete for \FO; and the results on \emph{conditional XPath} by Marx~\cite{Marx2005:conditional}, which imply that (a fragment of) \TRA is expressively complete for \FO over finite sibling-ordered trees.
Both are concerned with definability of binary relations.
Note however, that our objective differs from that of \cite{Ven90,Marx2005:conditional}: we are not restricted to linear orders or trees, and we are not primarily interested in expressive completeness with respect to \FO, but rather expressive completeness with respect to (semantic fragments of) Tarski's relation algebra, or, equivalently, $\FO^3$.

\paragraph{Acknowledgements}
Many thanks to Johan van Benthem, Ian Hodkinson, Luca Reggio,
Dimitri Surinx, and Evgenia Ternovska for helpful discussions and pointers. 
Balder ten Cate was supported by the European Union's Horizon
2020 research and innovation programme under grant MSCA-101031081. Brett McLean was supported by SNSF--FWO Lead Agen\-cy Grant 200021L\_196176/G0E2121N and by FWO Senior Postdoctoral Fellowship 1280024N.

\section{Preliminaries}

\paragraph{First-order logic and guarded second-order logic}

We restrict to structures over signatures consisting of binary relation symbols only. We write $\FO$ for first-order logic,
and we denote by $\FO^k$ (for $k\geq 1$) the $k$-variable
fragment of \FO, that is, the fragment of \FO consisting
of formulas that use only $k$ variables, where 
nested quantifiers may reuse the same variable.

We will also consider \emph{guarded second-order logic} (\GSO\
\cite{cohi}, also known as MSO$_2$ \cite{courcelle_book}),
which extends first-order logic with monadic second-order quantification
(i.e., quantification over sets) as well as guarded second-order quantification, by which we mean quantification over subrelations of
relations in the signature. Thus, for example, we can express in \GSO that a 
pair $(x,y)$ lies on a Hamiltonian cycle in a digraph, which is a property that cannot be 
expressed in MSO~\cite{Libkin2004:elements}.

By the \emph{quantifier rank} of a \GSO-formula $\phi$ we will mean the maximum nesting depth of first-order and/or second-order quantifiers.  We will write $A \equiv_{\GSO}^n B$ 
to indicate that two structures agree on all \GSO-sentences 
of quantifier rank at most $n$.

\paragraph{Binary relation algebras}

An \emph{$n$-ary operation on binary relations} is a map $O$ from 
first-order structures $A=(\dom(A), R_1^A, \ldots, R_n^A)$
to binary relations  $O(A)\subseteq \dom(A)^2$ that is 
isomorphism invariant: for every isomorphism
$h:A\cong B$, it holds that $h:O(A)\cong O(B)$.
Equivalently, one may think of an $n$-ary
operation on binary relations as mapping
first-order structures $A=(\dom(A), R_1^A, \ldots, R_n^A)$ to first-order structures
$A'=(\dom(A),O(A))$, where the domain of the structure
remains unchanged. 
We say that $O$ is
\emph{\FO-definable} if there is an \FO-formula $\phi(x,y)$
such that $O(A) = \{(a,b)\in \dom(A)^2\mid A\models\phi(a,b)\}$
for all $A$.
A \emph{binary relation algebra} is given by a collection $\mathcal{O}$ of
operations on binary relations.
We denote it by $\BRA(\mathcal{O})$.
We say that the algebra is FO if all its operations are
FO-definable.

\paragraph{Terms, term definable, finitely generated} 
Let $\mathbb{A}=\BRA(\mathcal{O})$ be  a binary relation algebra, and fix some countable infinite set of
binary relation symbols $R_1, R_2, \ldots$
  By an $n$-ary \emph{term} of $\mathbb{A}$ we mean a syntactic expression built up
  from the relation symbols $R_1, \ldots, R_n$ using the operations in $\mathcal{O}$ as function symbols.
  For instance, $R_1\cup R_1^\smile$ is an example of a 1-ary 
  \TRA-term. 
  We denote by $O_t$ the $n$-ary operation on binary relations defined by the term $t$.
We say that two $n$-ary terms $t$ and $t'$ are \emph{equivalent (in the finite)} if, for all (finite) structures
$A=(\dom(A),R_1^A,\ldots, R_n^A)$, $O_t(A)=O_{t'}(A)$. 
  We say that an operation on binary relations is \emph{term definable (in the finite)}
  in $\mathbb{A}$ if there is a term of $\mathbb{A}$ that defines it (over finite structures).
Note that, if $\mathcal{O}$ consists of \FO-definable operations, then 
every term of $\BRA(\mathcal{O})$ defines an \FO-definable operation. 
In fact, if every operation in $\mathcal{O}$ is $\FO^k$-definable (for some $k\geq 2$) then every $\BRA(\mathcal{O})$-term also defines an
$\FO^k$-definable operation. The same applies in the finite.

We say that a binary relation algebra $\BRA(\mathcal{O})$ is 
\emph{finitely generated} if there is a finite
subset $\mathcal{O}'\subseteq \mathcal{O}$, such that every operation
in $\mathcal{O}$ is term definable in $\BRA(\mathcal{O}')$. 

\paragraph{Tarski's relation algebra}

Tarski's relation algebra (\TRA) is an example of an \FO
binary relation algebra. It can be defined as
$\TRA \coloneqq \BRA(\id, \emptyset, -, \cap, \circ, {^\smile})$.
All operations in Table~\ref{tab:operations} are term definable in \TRA.
The following two classic results on \TRA will
be relevant for us.

\begin{thmC}[{\cite[Section 3.9]{Tarski1987:formalization}}]
\label{thm:tra-fo3}
    Both in general and in the finite:
    an operation on binary relations is term definable in \TRA
    if and only if it is $\FO^3$-definable. 
\end{thmC}

\begin{thmC}[\cite{Tarski41:RA,Lowenheim1915}]
\label{thm:fo-not-fin-gen}
    Both in general and in the finite:
    the binary relation algebra consisting of all \FO-definable
    operations is not finitely generated. 
\end{thmC}

Theorem~\ref{thm:fo-not-fin-gen} in fact follows from Theorem~\ref{thm:tra-fo3} together with the well-known fact in (finite) model theory that \FO does not collapse to any of its finite variable fragments; cf.~also~\cite[Theorem 2.13]{Ven90}.

\emph{Kleene Algebra} is an example of a non-\FO binary relation algebra, which includes the (\GSO-definable)
reflexive transitive closure operation. We omit the definition,
as we will not study it in this paper.

\section{The homomorphism-safe fragment is finitely generated}
\label{sec:monotone}

Recall that a homomorphism $h:A\to B$ is a function from 
the domain of $A$ to the domain of $B$ that preserves 
structure, i.e.~such that $(a,b)\in R^A$ implies $(h(a),h(b))\in R^B$.
We say that an operation $O$ on binary relations is \emph{homomorphism safe} if,
for every homomorphism $h:A\to B$
and $(a,b)\in O(A)$, $(h(a),h(b))\in O(B)$.
Equivalently, $O$ is homomorphism safe if and only if
every homomorphism $h:A\to B$ is also a homomorphism
$h:(A,O(A))\to (B,O(B))$, where $(A,O(A))$ denotes the 
expansion of the structure $A$ with $O(A)$ as an 
additional relation, and similarly for $(B,O(B))$.
Thus, intuitively, one can
think of homomorphism-safe operations as 
\emph{homomorphism-preserving} operations.

As indicated in Table~\ref{tab:operations}, examples of homomorphism-safe operations are $\cup$, 
$\cap$, and $\circ$, but not $-$.

\begin{thm}\label{thm:main-monotone}
    Both in general and in the finite:
    a \TRA-term is homomorphism-safe if and only if it is equivalent
    to a $\BRA(\id,\emptyset,\top,\circ,\cup,\cap,{^\smile})$-term.
\end{thm}

\begin{proof}
    The right-to-left direction can be proved by 
    a straightforward induction. We will focus
    on the more interesting left-to-right direction.

    We will make use of recent results regarding homomorphism-preserved \FO-formulas \cite{Bova19:how}.
Formally, we say that an \FO-formula $\phi(x_1,\ldots, x_n)$ is \emph{homomorphism preserved}  
if 
for every homomorphism $h:A\to B$
and tuple $a_1, \ldots, a_n$, we have $A\models\phi(a_1, \ldots, a_n)$ implies $B\models\phi(h(a_1),\ldots, h(a_n))$.
A classic theorem in model theory (known as the \emph{homomorphism preservation theorem}) 
states that 
a first-order formula is homomorphism preserved if and only if 
it is equivalent to a positive-existential \FO-formula
(i.e., a formula built up from atomic formulas using
only existential quantification,  conjunction, and disjunction).
Rossman~\cite{rossman2008homomorphism} proved that this holds also in the finite.
Bova and Chen~\cite[Corollary 24]{Bova19:how} further refined this to 
finite-variable fragments (both on arbitrary structures and in the finite): they showed that every homomorphism-preserved $\FO^k$ formula
    is equivalent to a positive-existential $\FO^k$-formula.

Let us now proceed with the proof of our theorem.
By Theorem~\ref{thm:tra-fo3},
   it suffices to show that every $\FO^3$-formula
   $\phi(x_1,x_2)$ (with two free variables) that is homomorphism preserved
   can be translated to the \TRA fragment in question.
   Moreover, by the aforementioned results of Bova and Chen, we may assume that $\phi(x_1,x_2)$ is a positive-existential
   $\FO^3$-formula. We 
   inductively translate $\phi(x_1,x_2)$ to a term $t$ in the
   specified fragment of \TRA, such 
   that $(a,b)\in O_t(A)$ iff $A$ satisfies $\phi$
   under the assignment that maps $x_1$ and $x_2$
   to $a$ and $b$, respectively.
   The base cases are straightforward. In particular,
   $R(x_1,x_2)$ translates to $R$, $R(x_2,x_1)$ 
   translates to $R^\smile$, $R(x_1,x_1)$ translates
   to $(R\cap id)\circ \top$, $x=y$ translates to 
   $\id$, etc.
   Conjunction and disjunction 
   translate to $\cap$ and $\cup$, respectively 
      (note that, here we take advantage of the fact that our induction
   hypothesis was stated specifically for formulas
   $\phi(x_1,x_2)$).
   Therefore, only the case remains where
   $\phi(x_1,x_2)$ is of the form 
   $\exists y\psi(x_1,x_2,y)$. It is not hard to 
   see that $\psi$ must, in this case, be a positive Boolean combination of formulas 
   with at most two 
   free variables. That is, $\psi$ can be 
   written as a disjunction of conjunctions of
   formulas with at most two free variables. 
   Furthermore, we can pull the disjunction 
   out from under the existential quantifier,
   and deal with it separately. 
   Therefore, we can assume without loss of generality that $\psi$ is a conjunction 
   of formulas with two free variables. 
   By grouping the conjuncts appropriately,
   we can write $\psi$ as $\psi_1(x_1,x_2)\land \psi_2(x_1,y)\land \psi_3(x_2,y)$. 
   By the induction hypothesis, each of these conjuncts
   can be translated to a \TRA-term, say, $t_1, t_2, t_3$. We can then
   translate $\phi$ as $t_1\cap (t_2\circ t_3^\smile)$. 
\end{proof}

It is worth comparing Theorem~\ref{thm:main-monotone} to results
by B\"orner and P\"oschel \cite{Borner1991:clones}, which state
that the ``logical clone'' (which is defined as the binary
relation algebra consisting of all \FO-definable operations on
binary relations) as well as the ``positive clone'' (the binary
relation algebra consisting of all  operations on binary
relations definable by a positive-existential \FO-formula) over
any fixed finite structure are finitely generated.  By
Rossman~\cite{rossman2008homomorphism}, the operations that can
be defined by a positive-existential \FO-formula are precisely
the homomorphism-safe \FO-definable operations.  We see that
Theorem~\ref{thm:main-monotone} is incomparable to the results just
mentioned.  On the one hand, it is only concerned with
\TRA-term-definable operations. On the other hand, it states that
there there is a finite basis of operations from which all
homomorphism-safe \TRA-terms are term definable \emph{over all
(finite) structures}.

One may wonder whether the approach taken in the proof of Theorem~\ref{thm:main-monotone} could be
used to establish a \L{}os--Tarski-style theorem 
for \TRA, characterising the 
fragment of \TRA that is preserved by the
$\subseteq$ relation, where, by $A\subseteq B$,
we mean that $A$ is an induced substructure of $B$.
More precisely we say that a first-order formula $\phi(x_1, \ldots, x_n)$
is $\subseteq$-preserved if, whenever $A\subseteq B$
and $a_1, \ldots, a_n\in \dom(A)$ and 
$A\models\phi(a_1, \ldots, a_n)$, then $B\models\phi(a_1, \ldots, a_n)$.
The classic \L os--Tarski preservation theorem  states that, on unrestricted (i.e., possibly infinite) structures,  an \FO formula is $\subseteq$-preserved if and only if it is equivalent to an existential \FO-formula.  As it turns out, however,  the 
\L os--Tarski theorem fails for $\FO^3$. More precisely, 
it has been shown \cite[Lemma 1 and 2]{andreka2023note}
that 
    there is a $\FO^3$-sentence over a signature
    consisting of a single binary relation,
    that is $\subseteq$-preserved, but that is not equivalent, even over finite structures, to an existential $\FO^3$-sentence.%
    \footnote{The result in~\cite{andreka2023note} is stated in terms of preservation under taking induced substructures, and it talks about the universal fragment of \FO. It is, however, equivalent by a duality argument.}
This shows that the approach we used for the
homomorphism-safe fragment of \TRA will not
work for the $\subseteq$-preserved fragment, which we also refer to as the \emph{$\subseteq$-safe} fragment.
So, it leaves the following question open. 

\begin{question}\label{question:homomorphism-safe}
Is the $\subseteq$-safe fragment of \TRA finitely generated? 
\end{question}

\section{The function-preserving fragment is not finitely generated}
\label{sec:func-pres}
Let $O$ be an $n$-ary operation on binary relations.
We say that $O$ is \emph{function preserving} if the following holds for all structures $A=(\dom(A),R_1^A,\ldots,R_n^A)$:
if  each $R_i^A$ is a partial function on $\dom(A)$,
then $O(A)$ is a 
partial function on $\dom(A)$.
Similarly, we say that 
$O$ is \emph{total-function preserving} if the following holds for all structures  $A=(\dom(A),R_1^A,\ldots,R_n^A)$:
if each $R_i^A$ is a total function on $\dom(A)$,
then $O(A)$ is a
total function on $\dom(A)$.

As indicated in Table~\ref{tab:operations},
the following are function preserving:
$\id$, $\emptyset$, $\D$, $\R$, $\antidom$, $\cap$, $\setminus$, $\circ$, $\ltimes$, and $\sqcup$.
Let us call the binary relation algebra
consisting of these operations \emph{function algebra} (\FA).
\FA was described in~\cite{hirsch} as ``in an informal sense at least, the richest natural case'' of an algebra of partial functions. In the same paper, a finite
axiomatisability result was established for \FA (see also~\cite{mclean-thesis} for a systematic study of
algebras of partial functions).

Our main result in this section is the following theorem.
\begin{thm}\label{thm:main}
Let $\mathcal{O}$ be any finite set of function-preserving \GSO-definable operations on binary relations. Then
there is a function-preserving operation on binary relations $O$
 that is term definable in \TRA but not in $\BRA(\mathcal{O})$, even over finite structures in which all relations are partial functions.
 \end{thm}

The proof will make use of the 
following lemma (where
$\uplus$ denotes the operation 
of \emph{disjoint union}).

\begin{lem}\label{lem:FV}
    For all structures $A, A', B, B'$ and 
    $n>0$, if $A\equiv_{\GSO}^n A'$ 
    and $B\equiv_{\GSO}^n B'$ then
    $A\uplus B\equiv_{\GSO}^n A'\uplus B'$.
\end{lem}

\begin{proof} 
The lemma can be derived from a (suitable adaptation to GSO of a) more general Feferman--Vaught theorem for MSO
\cite{Makowski2004:algorithmic}. However, here, we give
a direct argument using an Ehrenfeucht--Fraisse-style game argument. The game we will consider is played, 
as usual,
between two structures, $C$ and $C'$,
and has two players, Spoiler and Duplicator. 
In each round, Spoiler plays first and can make two types of moves: those
corresponding to first-order quantification and those corresponding 
to monadic or guarded second-order quantification. A move of the first type 
means that Spoiler picks an element of $C$ or of $C'$.
In this case, Duplicator must respond by picking a 
corresponding element of the other structure.
A move of the second type means that Spoiler picks
either a subset of the domain of $C$ or $C'$ 
(in which case Duplicator responds by picking a corresponding
subset of the domain of the other structure)
or a subrelation of one of the relations of $C$ or $C'$
(in which case, Duplicator responds by picking a 
coresponding subrelation of the same relation in the other structure). The game then continues using the same
pair of structures expanded with the chosen elements/sets/relations. The game is played for a 
fixed number of rounds, $n$. Duplicator wins if, 
after $n$ rounds, the resulting substructures
satisfy the same quantifier-free FO formulas
(with the chosen elements as parameters and chosen
sets/relations as relations). It is a standard exercise to show that Duplicator has a winning strategy for the $n$-round game if and only if $C\equiv^n_{\GSO} C'$.

We now use the above game to prove the statement.
Since $A\equiv_{\GSO}^n A'$ 
    and $B\equiv_{\GSO}^n B'$, Duplicator has 
    winning strategies in the two corresponding $n$-round games. Consider now the $n$-round game
    between $A\uplus B$ and $A'\uplus B'$.
    We will refer to $A$ and $B$ as the ``left'' part
    and the ``right'' part of $A\uplus B$ and 
    similarly for $A'$ and $B'$.
    Recall the two types of moves Spoiler can make in the game.
A move of the first type consists of choosing an element, which must belong either 
the ``left half'' of the structure or to the ``right
half''. In this case, Duplicator can respond using 
their assumed strategy in the corresponding game.
A move of the second type involves selecting either a set of elements, or a set of a tuples from a relation in the structure. In either case, the set in question
can be naturally partitioned into two halves, the ``left half'' and the ``right half''. Duplicator can therefore respond to each type of move simply by using her winning strategies for the two parts of the structure. It is easy to see that this yields a winning strategy for Duplicator.
\end{proof}

\begin{proof}[Proof of Theorem~\ref{thm:main}]
Let $n$ be a number greater than the maximum quantifier rank of the GSO-formulas defining 
the operations in $\mathcal{O}$.

For $m\geq 0$, let $C_m$ be the
directed graph that has a vertex
$a_{i,j}$ for every $i\in\{1, \ldots, m\}$ and $j\in\{1,2,3\}$, and that has an edge from $a_{i,j}$ to $a_{i',j'}$
  whenever $i'= (i \bmod m) +1$.
  In other words, $C_m$ is a directed cycle of length $m$ in which every vertex is replaced by three vertices. Then
let $C_m^\vee$ be the structure over the 
signature $\{f,g\}$ obtained from $C_m$
by replacing every edge by an $(f^\smile\circ g)$-path (using a fresh intermediate vertex each time). 
See Figure~\ref{fig:C}. We will refer to the vertices of the form $a_{i,j}$ as 
``normal nodes'' and the added intermediate vertices as ``auxiliary nodes''. In addition,
by a ``cluster of auxiliary nodes'' we mean the family of nine auxiliary nodes added between the points $a_{i,j}$ and $a_{(i+1 \bmod m),j'}$ for $j,j' \in \{1,2,3\}$, for some $i\in\{1, \ldots, m\}$.

\medskip\par\noindent\textbf{Claim 1:}
There are $m\neq m'$ such that,
in the structure 
$C := C_m^\vee\uplus C_{m'}^{\vee}$, 
all normal nodes satisfy the same
$\GSO$-formulas $\phi(x)$ of quantifier depth $n$ and likewise for the auxiliary nodes.

\medskip \emph{Proof of Claim 1:}
Since there are (up to equivalence) only finitely many 
\GSO-sentences of quantifier rank at most $n+1$,
by the pigeonhole principle, there
exist $m\neq m'$ such that $C_{m}^\vee\equiv_{\GSO}^{n+1} C_{m'}^\vee$.
Therefore, by Lemma~\ref{lem:FV},
$C_{m}^\vee\uplus C_{m'}^\vee \equiv_{\GSO}^{n+1} C_{m}^\vee\uplus C_{m}^\vee$. It follows by invariance 
under isomorphism that every 
normal node in $C_{m}^\vee\uplus C_{m}^\vee$ satisfies the same
\GSO-formulas $\phi(x)$, and
similarly for the auxiliary nodes. 
In other words, for all \GSO-formulas
$\phi(x)$, we have that
\[C_{m}^\vee\uplus C_{m}^\vee\models \forall x (\text{normal}(x)\to\phi(x))\lor\forall x(\text{normal}(x)\to\neg\phi(x))\] and 
\[C_{m}^\vee\uplus C_{m}^\vee\models \forall x (\text{auxiliary}(x)\to\phi(x))\lor\forall x(\text{auxiliary}(x)\to\neg\phi(x))\]
where $\text{normal}(x)$ is a shorthand
for $\exists yf(y,x)$ and 
$\text{auxiliary}(x)$ is a shorthand for 
$\exists yf(x,y)$. Since $C_{m}^\vee\uplus C_{m'}^\vee \equiv_{\GSO}^{n+1} C_{m}^\vee\uplus C_{m}^\vee$, the same holds in the 
structure $C_{m}^\vee\uplus C_{m'}^\vee$ for $\phi$
of quantifier rank at most $n$. 
This concludes the proof of Claim 1.

\medskip

\begin{figure}[t]
    \centering
\includegraphics[scale=1.2]{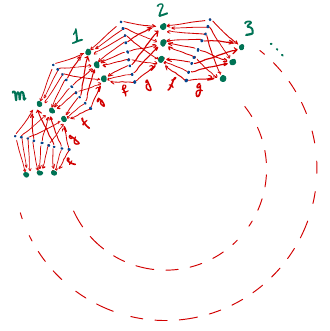}
\caption{Structure $C_m^\vee$}
    \label{fig:C}
\end{figure}

Note that the signature of $C$ is $\{f,g\}$ and that $f$ and $g$ are partial functions. 
Let $X$ be the set consisting of the following partial functions over the 
domain of $C$:
\begin{itemize}
    \item $f$,
    \item $g$,
    \item the identity function $\id$,
    \item $\id_1$ which is $\id$ restricted to the auxiliary nodes,
    \item $\id_2$ which is $\id$ restricted to the normal nodes,
    \item $f \cup \id_2$,
    \item $g \cup \id_2$,
    \item the empty relation $\emptyset$.
\end{itemize}
Each of the partial functions in $X$ is \TRA-term definable in $C$, and it will
be convenient to expand $C$ with these partial functions. That is,
we will treat $C$ as a structure over a signature consisting of these eight partial functions.

\medskip\par\noindent\textbf{Claim 2:}
Let $\phi(x,y)$ be any \GSO-formula  that is function-preserving. Then 
$C \models \phi(a,b)$ implies that $(a,b)$ belongs to
$f \cup g \cup \id$. In other words, $\phi$ defines a subrelation of
$f\cup g\cup \id$ in $C$. 

\medskip \emph{Proof of Claim 2:}
This can be shown using an
automorphism argument: suppose that 
$C\models\phi(a,b)$, and suppose, for the sake
of a contradiction, that
$b$ is not equal to $f(a)$, $g(a)$, or $a$ itself.
We will show that, then, there exists
some $b'\neq b$ such that $(C,a,b)\cong (C,a,b')$, 
and therefore $C\models\phi(a,b')$, 
contradicting the assumption
that $\phi(x,y)$ was function preserving. 
We argue by cases. 
First, suppose that $a$ is a normal node.
We may assume without loss of
generality that $a = a_{1,1}$.
Recall that $(a,b)\not\in \id$.
If $b=a_{1,2}$ or $b=a_{1,3}$, then we can pick $b'$ to be $a_{1,3}$, respectively, $a_{1,2}$.
It then follows from the construction
of the structure $C$ that  $(C,a,b)\cong (C,a,b')$.
Similarly, if $b=a_{i,j}$ with $i\neq 1$,
then it follows from the construction
of the structure $C$ that $(C,a,b)\cong (C,a,b')$
for all $b'=a_{i,j'}$.
Finally,  if $b$ is an auxiliary node, then 
it follows from the construction
of the structure $C$ that $(C,a,b)\cong (C,a,b')$
for some  auxiliary node $b'\neq b$ from the same cluster. 
This concludes the case where 
$a$ is a normal node. Next, suppose that $a$ is an 
auxiliary node, and recall that $(a,b)\not\in f\cup g\cup \id$. Regardless whether $b$ is a normal or a special node, it follows that $a$ and $b$ do not co-occur in any fact (i.e., tuple in a relation) of $C$. It easy to see that, then,
$(C,a,b)\cong (C,a,b')$ must be satisfied if we choose $b'\neq b$ to be another node from the same cluster as $b$. This concludes the proof of Claim 2.

\medskip\par\noindent\textbf{Claim 3:}
Let $\phi(x,y)$ be any \GSO-formula of quantifier rank less than $n$ that is function-preserving. If $C \models \phi(a,b)$ and $f(a)=b$, then for all $a'$ and $b'$
with $f(a')=b'$ we have that $C \models \phi(a',b')$. Likewise
for the functions $g$, $\id_1$, and $\id_2$.

\medskip \emph{Proof of Claim 3:}
We will discuss the proof for the case for $f$.
The same argument applies to $g$, 
while the cases for $\id_1$ and $\id_2$ 
follow immediately from Claim~1.
Assume $C,a,b\models\phi(x,y)$. Then $C,a\models\exists y(f(x,y)\land \phi(x,y))$. Therefore, by Claim~1, we have $C,a'\models\exists y(f(x,y)\land\phi(x,y))$, and
therefore, since $f$ is a partial function and $f(a')=b'$, we have $C,a',b'\models\phi(x,y)$.
This concludes the proof of Claim~3.

The next claim follows  from Claim 2 and 3.

\medskip\par\noindent
\textbf{Claim 4:} If $\phi(x,y)$ is any \GSO-formula of quantifier rank less than $n$ that is function-preserving, then 
the partial function defined by $\phi(x,y)$ in $C$ belongs to $X$.

\medskip \emph{Proof of Claim 4:}
By Claim 2,
the relation $R=\{(c,d)\mid C,c,d\models\phi(x,y)\}$
is contained in $f\cup g\cup \id_1\cup \id_2$, while
by Claim 3, $R\cap f\neq\emptyset$ implies
$f\subseteq R$, and likewise for $g$, $\id_1$ and
$\id_2$. It follows that $R$ must be equal to the union
of a subset of the relations $f, g, \id_1, \id_2$. In other words,
$R$ belongs to $X$. 
This concludes the proof of Claim~4.

Claim 4 tell us that no function-preserving \GSO-operation with quantifier rank smaller than $n$ can take us
outside of the set $X$. Since each operation in $\mathcal{O}$ is defined
by a \GSO-formula of quantifier rank less
than $n$, and is function preserving, this implies, by induction, that
every term  of $\BRA(\mathcal{O})$ denotes one of the relations
in $X$ in $C$. 

This implies the theorem:
consider the \TRA-term 
$(f^\smile\circ g)^{m}\cap \id$,
 where $(\,\cdot\,)^m$ stands for an $m$-fold composition.
This term denotes the identity relation
restricted to the normal nodes of $C_m$ only;
this relation does not belong to $X$. Therefore, this term cannot
be equivalent to any term of $\BRA(\mathcal{O})$. Nevertheless
it is function preserving, simply because its interpretation always 
consists only of reflexive edges.
\end{proof}

With some minor modifications, the same argument applies to total-function-preserving operations:

\begin{thm}\label{thm:main-total}
Let $\mathcal{O}$ be a finite set of total-function-preserving \GSO-definable operations on binary relations. Then
there is a total-function-preserving operation $O$
 that is term definable in \TRA but not in $\BRA(\mathcal{O})$, 
 even over finite structures in which
every relation is a total function.
\end{thm}

\begin{proof} (sketch)
We use the same construction as before, except that we extend the structure $C$
with an additional ``sink node'' $s$ and an additional function $\hat\emptyset$ where $\hat\emptyset(c)=s$
for all nodes $c$ (including $s$ itself). Observe that $\hat\emptyset$ is a total
function. We also extend the partial functions $f$ and $g$ to total functions $\hat{f}$ and
$\hat{g}$, by
setting $\hat{f}(c)=\hat{g}(c)=s$ for every normal node $c$ and $\hat{f}(s)=\hat{g}(s)=s$. 
Note that the old partial functions $f$ and $g$ are \TRA-term definable from the new 
ones, namely as $f = \hat{f}- (\top \circ \hat\emptyset)$ and $g = \hat{g} - (\top \circ \hat\emptyset)$.
Now the same argument as before shows that the \TRA-term
\[ \big( (f^\smile\circ g)^{m}\cap \id\big)\sqcup \hat\emptyset\]
(where $f$ and $g$ are now shorthand for the aforementioned terms, and where $\sqcup$ is the preferential union operator)
defines a total-function-preserving operation that 
is not term definable in $\BRA(\mathcal{O})$.
\end{proof}

As a consequence of Theorem~\ref{thm:main}, we obtain the following.

\begin{cor} Both in general and in the finite:
\begin{enumerate}
\item The function-preserving fragment of\/ \TRA is not finitely generated. In particular, not every function-preserving \TRA-term is term definable in \FA.
\item The homomorphism-safe function-preserving fragment of\/ \TRA is not finitely generated.
\item The $\subseteq$-safe function-preserving fragment of\/ \TRA  is not finitely generated.
\end{enumerate}
\end{cor}

\begin{proof}
The first item follows immediately from Theorem~\ref{thm:main}. 
The other items follow from its proof.
This is because the \TRA-term used as counterexample in the proof, i.e., 
$(f^\smile\circ g)^{m}\cap \id$, uses only operations that are homomorphism safe and  $\subseteq$-safe.
(Note that the same does \emph{not} hold in the total-function-preserving case because there we used preferential union.)
\end{proof}

\begin{question}
    Is the homomorphism-safe total-function-preserving fragment of \TRA finitely generated?
\end{question}

Given that the function-preserving fragment of \TRA is not finitely generated, one may ask if it is at least generated by a \emph{recursive} set of operations. This is 
indeed the case, for a trivial reason: 
for any \TRA term $t$, consider the term
$t'=t\setminus(t\circ (\top\setminus \id))$. 
By construction $t'$ always outputs a partial
function. Furthermore, on any input where $t$
produces a partial function, $t'$ produces the 
same output as $t$. Therefore, the function-preserving
fragment of \TRA is generated by the (recursive)
set of all \TRA-terms of the form $t\setminus(t\circ (\top\setminus \id))$.

Another question left open by the above results is
whether $\FA$, although it is not the function-preserving
fragment of $\TRA$, can still be characterized as a 
natural fragment of $\TRA$. 

\begin{question} Can \FA be characterised as a fragment of \TRA using additional properties besides function preserving (or using a strengthening of the notion of ``function preserving'')?
\end{question}

\section{The forward function-preserving fragment is finitely generated}
\label{sec:func-pres-positive}

In our proof of Theorem~\ref{thm:main}, 
we implicitly made use of 
the fact that any binary relation can be 
represented as a composition $f^\smile\circ g$,
where $f,g$ are partial functions. That is, 
we crucially made use of the inverse operation.
This is indeed essential to the proof: as we will 
now show, if we 
restrict attention to direction-preserving 
operations (\emph{forward} operations, as we will call them below), then we do get a binary relation algebra
that is finitely generated. 

Formally, we say that an $n$-ary operation $O$ on binary relations is \emph{forward} if 
    for all structures $A$ over signature $\sigma=\{R_1, \ldots, R_n\}$ and for all pairs $(a,b)\in \dom(A)$, we have that
      $(a,b)\in O(A)$ if and only if $(a,b)\in O({A_a})$ where $A_a$ is the substructure of $A$ generated by $a$, i.e., 
      the induced substructure of $A$ whose domain consists of
      all elements reachable from $a$ by a finite directed path along the relations $R_1^A, \ldots, R_n^A$.
      In particular, this implies that, whenever $(a,b)\in O(A)$ then $b$ must belong to $A_a$. We say that 
      $O$ is \emph{forward over a class of structures $K$} if 
      the above holds for all structures $A\in K$.%
      \footnote{Note that being forward is a stronger 
      requirement than requiring that $b\in A_a$ for all $(a,b)\in O(A)$. Indeed, the operation defined by the \TRA-expression
      $R\cap (R^\smile\circ R)$ satisfies the latter requirement but is not forward.}

\begin{lem}\label{lem:bounded-depth}
Let $K$ be any FO-definable class of structures, and 
let $O$ be any FO-definable operation on binary relations that is forward over $K$. Then there is a natural number $m$ such that,
for all structures $A\in K$ and $a,b\in \dom(A)$, whether
$(a,b)$ belongs to $O(A)$ depends only on the substructure
of $A$ consisting of the elements reachable from $a$ by a 
directed path of length at most $m$.
\end{lem}

\begin{proof}
    This can be shown using a simple compactness argument
    \cite{vanBenthem2007:new}:
    let $\chi$ be the FO-sentence defining $K$, and 
    let $n$ be the arity of the operation $O$. By assumption,
    $O$ is defined by a first-order formula $\phi(x,y)$ 
    over the signature consisting of the relation symbols 
    $R_1, \ldots, R_n$. Let $P$ be a fresh unary relation symbol,
    let $\phi^P$ be the result of relativising all quantifiers
    in $\phi$ by $P$ (i.e., replacing $\exists z$ by $\exists z(P(z)\land\ldots)$ and replacing $\forall z$ by $\forall z(P(z)\to\ldots)$). Furthermore, for every natural number $k$, let 
    $\psi_k(x)$ be the \FO-formula expressing that 
    all elements reachable from $x$ by a directed path of length at most $k$ satisfy $P$. 
    Then $\{\chi, \psi_k(x)\mid k\geq 0\}\models \forall y(\phi(x,y)\leftrightarrow (P(y)\land \phi^P(x,y)))$.
    It follows by compactness that, for some $m$,
    $\{\chi, \psi_k(x)\mid 0\leq k\leq m\}\models \forall y(\phi(x,y)\leftrightarrow (P(y)\land \phi^P(x,y)))$.
    This proves the lemma.
\end{proof}

\begin{thm}\label{thm:forward-function}
    Let $K_\textup{pf}$ be the class of structures in which each relation is a partial function, and let $O$ be any FO operation on binary relations. The following are equivalent:
    \begin{enumerate}
        \item $O$ is function preserving and forward over $K_\textup{pf}$, 
        \item 
    $O$ is term-definable in $\BRA(\circ,\antidom, \cap, \sqcup)$
    over $K_\textup{pf}$.
    \end{enumerate}
\end{thm}

\begin{proof}
    The direction from 2 to 1 is straightforward. For the direction from 1 to 2:
    let $O$ be any $n$-ary \FO operation that is function preserving and forward over $K_\textup{pf}$. From the fact that 
    $O$ is forward over $K_\textup{pf}$, it follows by Lemma~\ref{lem:bounded-depth}
    that there exists a constant $m>0$ (depending on $O$) such 
    that whether a pair $(a,b)$ belongs to $O(A)$, for $A\in K_\textup{pf}$,
    depends only on the substructure $B\subseteq A$
    consisting of the elements reachable from $a$
    by a directed path of length at most $m$. 
    For $A\in K_\textup{pf}$,
    such a substructure $B$ can be of size at most
    $(n+1)^m$. 
    There are only finitely
    many isomorphism types of such structures $B$. 
    Furthermore, for each such $B$, the structure 
    $(B,a)$ can be characterised up to isomorphism
    by an intersection $\chi_{B,a}$ of terms of the following
    forms:
    \begin{itemize}
        \item ${\sim} (f_1\circ\cdots\circ f_k)$ \\ ``there is no outgoing $f_1\circ\cdots\circ f_k$ path''
        \item ${\sim\sim} (f_1\circ\cdots\circ f_k)$ \\ ``there is an outgoing $f_1\circ\cdots\circ f_k$ path''
        \item ${\sim} (f_1\circ\cdots\circ f_k\cap g_1\circ\cdots\circ g_l)$ \\ ``the outgoing $f_1\circ\cdots\circ f_k$ path and the outgoing $g_1\circ\cdots\circ g_l$ path do not lead to the same node''
        \item ${\sim\sim} (f_1\circ\cdots\circ f_k\cap g_1\circ\cdots\circ g_l)$ \\ ``the outgoing $f_1\circ\cdots\circ f_k$ path and the outgoing $g_1\circ\cdots\circ g_l$ path do  lead to the same node''
    \end{itemize}
    Note that here we implicitly use $\id$ (which is definable as ${\sim}({\sim} f \circ f)$)  for the case where $k=0$ or $l=0$. 
    Finally, we can take our term to be $\chi_{B,a}\circ (f_1\circ\cdots\circ f_k)$
    where $f_1, \ldots, f_k$ describes an arbitrary
    directed path from $a$ to $b$ (or simply $\chi_{B,a}$ if the path is empty). 
    Doing this for each isomorphism type of structure $B\models\phi(a,b)$, we obtain finitely many terms (defining relations that are guaranteed to be pairwise disjoint from each other) we then 
    combine using the preferential union operator 
    (in arbitrary order, since they are pairwise disjoint). In the special case where there
    is no $B\models\phi(a,b)$, we may choose as our 
    term $\emptyset$ (which is definable as ${\sim}f \circ f$).
\end{proof}

The collection $\{\circ,\antidom, \cap, \sqcup\}$ of operations
identified in Theorem~\ref{thm:forward-function} is one that has
already been investigated in the literature. Specifically,
Jackson and Stokes
\cite{DBLP:journals/ijac/JacksonS11} give a finite equational
axiomatisation of the class of algebras isomorphic to a set of
partial functions equipped with these operations.
The equational theory of these algebras is coNP-complete \cite{hirsch}.

\begin{question}
Does Theorem~\ref{thm:forward-function} hold in the finite?
\end{question}

Although we do not know the answer to this question, we can show that Lemma~\ref{lem:bounded-depth} fails in the finite, and therefore, a different approach is required.

\begin{prop}
    \label{prop:bounded-finite-failure}
    Lemma~\ref{lem:bounded-depth} fails when $K$ is the class of all finite structures (which is not FO-definable).
\end{prop}

\begin{proof}
 Let $\phi(u)$ be the conjunction of the following \FO-formulas:
  \begin{itemize}
    \item $R_3(u,u)$
    \item $\forall v(R_3(u,v)\to \exists w(R_3(v,w)\land R_2(w,v)))$
    \item $\forall vw(R_3(u,v)\land R_3(v,w)\to R_3(u,w))$
    \item $\neg\exists v R_2(u,v)$
    \item $\forall vw(R_3(u,v)\land (\exists^{\geq 2}s R_2(v,s))\land R_1(u,w)\to R_4(w,v))$
\end{itemize}
It follows from the fact that every quantifier is bounded by a forward-oriented atom, that $\phi(u)$ is invariant for generated
substructures~\cite{Feferman1968Persistent}. That is, for all structures $A$ and elements $a$, we have
$A\models\phi(a)$ if and only if $A_a\models\phi(a)$.

Next, let $\psi(x,y) := (x=y)\land \exists u(R_1(u,x)\land\phi(u))$.

It follows immediately from the presence of the equality conjunct that $\psi(x,y)$ is function preserving. 

\medskip\par\noindent\textbf{Claim 1:} 
$\psi(x,y)$ defines a forward operation, i.e.,  
for all 
finite structures $A$, we have $A\models\psi(a,a)$ if and only if 
$A_a\models\psi(a,a)$. 

\medskip
The right-to-left direction is easy (and does not depend on the restriction to finite structures). For the other direction,
suppose that $A\models\psi(a,a)$. It follows, by the construction of $\psi$ and the finiteness of the structure $A$, that there exist elements connected as in Figure~\ref{fig:counterexample-forward}.
(In fact, further facts hold that have not been drawn
in the figure to avoid cluttering. Specifically,
$R_3(b_i,b_j)$ holds for all $i< j$.)

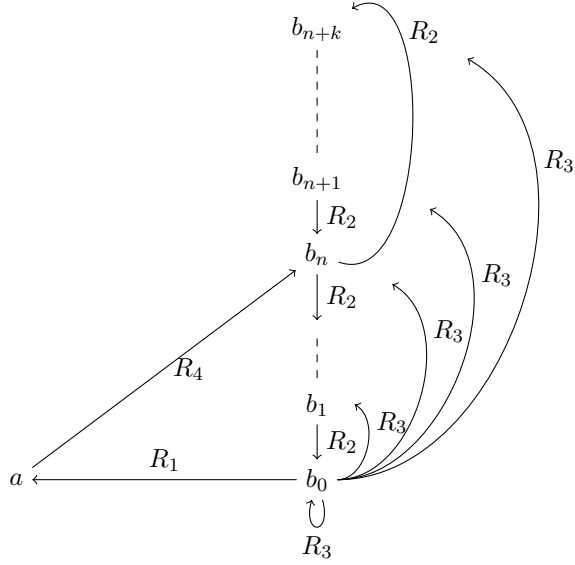
\begin{figure}
\begin{center}
\begin{tikzpicture}
  [align=center,scale=1, every node/.style={scale=.9}]
    \node (a) at (-1,0) {$a$};
    \node (b0) at (3,0) {$b_0$};
    \node (b1) at (3,1) {$b_1$};
    \node (bn0) at (3,2) {};
    \node (bn) at (3,3) {$b_n$};
    \node (bnn) at (3,4) {$b_{n+1}$};
    \node (bm) at (3,6) {$b_{n+k}$};
    \draw [-to] (b0) -- node[above] {$R_1$} (a);
    \draw [-to] (b1) -- node[right] {$R_2$} (b0);
    \draw [-to] (bn) -- node[right] {$R_2$} (bn0);
    \draw [-to] (bnn) -- node[right] {$R_2$} (bn);
    \draw [dashed] (bn0) -- (b1);
    \draw [dashed] (bm) -- (bnn);
    \draw [-to] (bn) to [out=340,in=30] node[right, near end] {$R_2$} (bm);
    \draw [-to] (b0) to [out=0,in=330] node[right, near end] {$R_3$} (3.5,1);
    \draw [-to] (b0) to [out=0,in=330] node[right, near end] {$R_3$} (4,2.6);
    \draw [-to] (b0) to [out=0,in=330] node[right, near end] {$R_3$} (4.5,3.6);
    \draw [-to] (b0) to [out=0,in=330] node[right, near end] {$R_3$} (5,5.6);
    \draw [-to] (b0) to [loop below] node[below] {$R_3$} (b0);
    \draw [-to] (a) to node[right] {$R_4$} (bn);
\end{tikzpicture}
\end{center}
\caption{Structure satisfying $\psi(a,a)$.}
\label{fig:counterexample-forward}
\end{figure}

It follows that all the depicted elements belong to $A_a$.
In particular, $b_0$ belongs to $A_a$.
From this, it follows that $A_a\models\psi(a,a)$.
This concludes the proof of Claim 1.

Now let $m$ be any natural number. 
Let $A$ be the structure drawn above, with $n=m+1$.
Let $B$ be the identical structure but with the node $b_0$
removed. Clearly, $A\models\psi(a,a)$ and 
$B\not\models\psi(a,a)$ (because $B$ lacks a reflexive $R_3$-edge). However, the induced substructures consisting of 
nodes reachable from $a$ by a directed path of length at most $m$ 
are identical.
\end{proof}

Proposition~\ref{prop:bounded-finite-failure}, incidentally, also
resolves in the
negative an open question about hybrid logic posed in \cite[Section 7]{Abramsky2022:comonadic},
namely whether the technique used
\cite{Abramsky2022:comonadic} for proving a preservation theorem
for hybrid temporal logic in the finite could be extended to
prove a similar result for the case without backward modalities. It follows from Proposition~\ref{prop:bounded-finite-failure}
that the corresponding preservation theorem in the 
finite in fact fails for hybrid logic without backward modalities.

\medskip

We can adapt the proof of Theorem~\ref{thm:forward-function} to obtain a similar, but undirected, result for \emph{injective} partial functions. For this, we say that $O$ is \emph{injective-function preserving} if the following holds for all structures $A=(\dom(A),R_1^A,\ldots,R_n^A)$:
if  each $R_i^A$ is an injective partial function on $\dom(A)$,
then $O(A)$ is an 
injective partial function on $\dom(A)$. Let us also say that that an $n$-ary operation $O$ on binary relations is \emph{local} if 
    for all structures $A$ over signature $\sigma=\{R_1, \ldots, R_n\}$ and for all pairs $(a,b)\in \dom(A)$, we have that
      $(a,b)\in O(A)$ if and only if $(a,b)\in O({A^\leftrightarrow_a})$ where $A^\leftrightarrow_a$ is 
      the induced substructure of $A$ whose domain consists of
      all elements reachable from $a$ by a finite undirected path along the relations $R_1^A, \ldots, R_n^A$.
      As before, this implies that, whenever $(a,b)\in O(A)$ then $b$ must belong to $A^\leftrightarrow_a$.

      To state the result, we first define a variant of preferential union that is injective-function preserving. We call this new operation \emph{injective union} and use $\iunion$ to denote it. The operation adds to its first argument any pairs from its second argument whose addition does not violate functionality or injectivity. One possible term definition of injective union is as follows.
      
      \[f\iunion g \coloneqq (f \sqcup g) \cap (f^\smile \sqcup g^\smile)^\smile\]

      \begin{thm}\label{thm:forward-backward-function}
      Let $K_\textup{ipf}$ be the class of structures in which each 
      relation is an injective partial function, and let $O$ be any FO operation on binary relations. The following are equivalent:
      \begin{enumerate}
          \item $O$ is injective-function preserving and local over~$K_\textup{ipf}$.
          \item $O$ is term-definable in $\BRA(\circ,\antidom, \cap, {^\smile}, \iunion)$ over $K_\textup{ipf}$.
      \end{enumerate}
\end{thm}

\begin{proof}(sketch) 
    First note that we can obtain an undirected analog of Lemma~\ref{lem:bounded-depth} using a similar proof. That is, if an
    FO-definable operation is local over $K_\textup{ipf}$, then there is a natural number $m$ such that, for all
    structures $A\in K_\textup{ipf}$ and $a,b\in\dom(A)$, 
    whether $(a,b)$ belongs to $O(A)$ depends only on the substructure of $A$ consisting of the elements reachable from $a$ by an undirected path of length at most $m$.

    Next, the same proof used for Theorem~\ref{thm:forward-function} works if we replace every instance of `directed path' by `oriented path' (i.e., sequence of possibly reverse-oriented edges), use $^\smile$ to express reverse-oriented edges in such paths, and use $\iunion$ in place of $\sqcup$.
\end{proof}

The collection $\{\circ,\antidom, \cap, {^\smile}, \iunion\}$ of
operations identified in
Theorem~\ref{thm:forward-backward-function} is one that has been
considered in the literature on inverse semigroups. Any set of
injective partial functions closed under these operations forms a
Boolean inverse monoid in the sense of Lawson \cite{Lawson_2010};
indeed these are the canonical examples of Boolean inverse
monoids.\footnote{In the definition of Boolean inverse monoids, only joins of pairs of \emph{orthogonal} elements are required, that is, elements $a$ and $b$ such that $a^\smile \circ b$ and $a \circ b^\smile$ are both zero. This is equivalent to the presence of $\iunion$, since clearly if a collection of injective partial functions is closed under $\iunion$ then it is closed under orthogonal joins, and conversely, in the presence of the other operations, $\iunion$ is expressible as the orthogonal join of $\antidom b \circ a \circ \antidom(b^\smile)$, $\antidom a \circ b \circ \antidom(a^\smile)$, and $a \cap b$.} Conversely, from the results of Lawson it can be seen that any Boolean inverse monoid is isomorphic to one of these algebras of injective partial functions \cite[Proposition~2.23(2)]{Lawson_2010}. Thus Theorem~\ref{thm:forward-backward-function} demonstrates that within the program of studying enrichments of inverse semigroups, the Boolean inverse monoids are in a sense the \emph{fully enriched} instances. 

\section{Conclusion}
\label{sec:conclusion}

In summary, our results show that certain semantic fragments of Tarski's relation algebra, such as the homomorphism-safe fragment, admit a syntactic characterisation in terms of a finite set of operations, while others, such as the function-preserving fragment, do not. We hope that these results show that the study of preservation theorems in
the context of algebras of binary relations is an
interesting topic.
We conclude by listing a few directions that deserve further exploration. 

Firstly, one could explore the same questions for other semantic properties of operations on binary relations (e.g., $\subseteq$-safety, as mentioned in Section~\ref{sec:monotone}, as well as additivity
\cite{additive}). 
Secondly, our results
concern fragments of \TRA, but the same
questions can be asked for
other binary relation algebras, including ones that contain the transitive closure operator.
In particular, our results leave open the question whether the function-preserving fragment of Kleene Algebra with Tests (KAT) is finitely generated. 

Finally, various applications of \TRA in computer science and elsewhere are concerned with a restricted class of structures, such as finite trees (e.g., XPath), linear orders (e.g., interval temporal logics), or 
variable-assignment spaces (e.g., dynamic predicate logic~\cite{dplogic} and the Logic of Information Flows (LIF)~\cite{lif_frocos,heba-thesis}). 
It is therefore meaningful to ask whether our results hold also over these restricted classes of structures. 

\bibliographystyle{alphaurl}
\bibliography{bib}

\end{document}